\documentstyle[12pt,emulateapj,epsfig]{article} 

\newcommand{\kms}{km s$^{-1}$}       
  
\newcommand{\cmthree}{cm{$^{-3}$}}

\newcommand{\um}{$\mu$m}                                 
 
\newcommand{\lsun}{$L_{\odot}$}               
 
\newcommand{\rsun}{$R_{\odot}$}

\newcommand{\lsim}{\;\lower.6ex\hbox{$\sim$}\kern-7.75pt\raise.65ex\hbox{$<$}\;} 
\newcommand{\gsim}{\;\lower.6ex\hbox{$\sim$}\kern-7.75pt\raise.65ex\hbox{$>$}\;} 
 
\newcommand{\amin}{$^{\prime}$}                   
\newcommand{\asec}{$^{\prime \prime}$}

\newcommand{\lflux}{W cm$^{-2}$} 
\newcommand{\fdens}{W cm$^{-2}$ \um\^{-1}} 
\newcommand{\oia}{[O{\sc i}]63\um} 
\newcommand{\oib}{[O{\sc i}]145\um} 
\newcommand{\oiiiv}{[O{\sc iii}]} 
\newcommand{\oiii}{[O{\sc iii}]88.3\um} 
\newcommand{\cii}{[C{\sc ii}]158\um}

\newcommand{\nii}{[N{\sc ii}]122\um}

\newcommand{\bo}{${\bf \vec{B}_0}$} 
 
\newcommand{\mbop}{$B_{0\perp}$} 
\newcommand{\vs}{$\vec{v}_s$}

\newcommand{\mug}{$\mu$G} 
\newcommand{\ha}{H$\alpha$} 
\newcommand{\oic}{[O{\sc i}]6300\AA}
\newcommand{\oiiib}{[O{\sc iii}]5007\AA}
\newcommand{\go}{G$_{\rm 0}$}

\begin{document} 
%
%
%
\lefthead{Molinari, Noriega-Crespo \& Spinoglio} 
\righthead{Shock-Induced PDR in the HH\,80/81 Flow} 
 
\title{A Shock-Induced PDR in the HH\,80/81 Flow. Far Infrared
Spectroscopy.$^0$} 
 
\author{Sergio Molinari}
\affil{Infrared Processing and Analysis Center, California  Institute of 
Technology, MS 100-22, Pasadena, CA 91125, USA} 
\affil{Electronic mail: molinari@ipac.caltech.edu} 

\author{Alberto Noriega-Crespo} 
\affil{SIRTF Science Center, California Institute of Technology} 
\affil{IPAC 100-22, Pasadena, California, 91125} 
\affil{Electronic mail: alberto@ipac.caltech.edu} 

\affil{\&}  

\author{Luigi Spinoglio} 
\affil{CNR-Istituto di Fisica dello Spazio Interplanetario, Area di  
Ricerca Tor Vergata, via Fosso del Cavaliere I-00133 Roma, Italy} 
\affil{Electronic mail: luigi@ifsi.rm.cnr.it} 

\submitted{Accepted 2000 August 24}
 
\footnotetext[0]{Based on observations with ISO, an ESA project with 
instruments funded by ESA Member States (especially the PI countries: 
France, Germany, the Netherlands and the United Kingdom) with the 
participation of ISAS and NASA.} 

\begin{abstract} 
 
The two spectrometers on board the Infrared Space Observatory were used
to observe the Herbig-Haro objects HH\,80, 81 and 80N, as well as their
candidate exciting source IRAS\,18162-2048. The fine structure lines of
\oia, \oib\ and \cii\ are detected everywhere, while \nii\ and \oiii\
are only detected toward the HH objects; line ratios confirm for the
first time the collisionally excited HH nature of HH\,80N. No molecular
line is detected in any of the observed positions. We use a full shock
code to diagnose shock velocities v$_s\sim$100 \kms\ toward the HH
objects, as expected from the optical spectroscopy. Since proper motions
suggest velocities in excess of 600 \kms, the HH objects probably
represent the interface between two flow components with velocity
differing by  $\sim$v$_s$. Aside from the flow exciting source, the
\cii\ line is everywhere brighter than the \oia\ line, indicating the
presence of a Photo-Dissociation Region (PDR) all along the flow.
Continuum emission from the HH objects and from other positions along
the flow is only detected longword of $\sim$50\um, and its
proportionality to the \cii\ line flux suggests it is PDR in origin. We
propose that the FUV continuum irradiated by the HH objects and the jet
is responsible for the generation of a PDR at the walls of the flow
cavity. We develop a very simple model which strengthens the
plausibility of this hypothesis.

\end{abstract} 
%
 
\section{Introduction} 
\label{intro} 
 
Stellar jets and outflows arising from low mass protostellar objects
are  considered ubiquitous, since essentially each protostar must go
through an active period of mass loss to get rid of its angular momentum
to become a star (c.f. Hartmann \cite{har98}). These outflows are well
collimated and  supersonic and interact mostly through shocks with the
interstellar medium. When these shock excited regions are detected by
optical means, they are  identified as Herbig-Haro objects (Reipurth \&
Heathcote \cite{RH97}).

The process of mass loss for intermediate and high mass protostellar
objects is less well understood (Churchwell \cite{chur98}).
Observationally  very few of these objects have collimated optical jets
or well defined molecular  outflows (Poetzel,  Mundt \& Ray
\cite{PMR89}; Shepherd \& Churchwell~\cite{SC96}). The Herbig-Haro 80-81
system, which is driven by a 20\,000 \lsun~IRAS source, is one of those
rare cases with a well collimated supersonic jet that reaches flow
velocities of $\sim 600 -1400$ \kms (Mart{\'{\i}} et al. \cite{MRR95},
\cite{MRR98}). HH\,80/81 are at the edge of the L291 cloud in
Sagittarius  (Reipurth \& Graham \cite{RG88}) at an estimated
kinematical distance of  1.7 kpc (Rodr{\'{\i}}guez et al.
\cite{rodri80}). At this distance the HH\,80/81  system, with an angular
size of  $\sim$10\amin.8, spans $\sim 5.3$ pc. The proper motion 
measurements confirm that IRAS\,18162-2048 is the  source driving the
outflow, which is part of a small cluster of infrared  sources  (Aspin
\& Geballe \cite{AG92}; Aspin et al. \cite{aspin94}). IRAS\,18162-2048,
given its large luminosity, is  likely to become a B type star. Most of
the early work on HH\,80/81 was performed at radio wavelengths, where
3.6 and 6 cm observations found the  collimated radio jet emanating 
from IRAS\,18162-2048, as well as a section of the  counter-flow, named
HH\,80N,  which is optically invisible (Rodr\'{\i}guez\&
Reipurth~\cite{RR89}; Mart\'{\i}, Rodr\'{\i}guez \&
Reipurth~\cite{MRR93}, hereafter MRR93).

Recently  Heathcote, Reipurth \& Raga (\cite{HRR98}, hereafter HRR) have
carried a detailed analysis of HH\,80/81 morphology and kinematics,
using HST WFPC2 images and ground based optical images and spectra. They
concluded in their study that the emission arising from HH\,80/81  is
due to shocks with velocities up 600 \kms. The energy release by such
shocks is so high that thermalizes the gas to temperatures $> 10^6$ K,
i.~e. the shocks are adiabatic.  If the HH objects were to resemble bow 
shock structures (Raga \& B\"ohm \cite{RB86};  Hartigan et al.
\cite{HRH87}), then the optical emission observed in \ha\ or \oiiiv~
$\lambda$5007 comes from the `wings', while the stagnation region (the
tip of the bow) is several  times further ahead and invisible. This
scenario, however, is derived by modeling the optical line profiles
(HHR98); from the morphological viewpoint, the HH objects appear as
irregular knots that do not resemble wings of a bow.

In this study we present ISO (Kessler et al.~\cite{Ketal96})
spectroscopic observations of the HH\,80/81/80N system, including their
exciting source  IRAS\,8162-2048. The observations and data analysis
are  presented in $\S$\ref{obs} and the results in $\S$\ref{res}. A
model to interpret the results is presented in $\S$\ref{model} and
discussed in $\S$\ref{discussion}. We summarize our conclusions in
$\S$\ref{conclusions}.

\section{Observations} 
\label{obs} 

\begin{table*}
\begin{center}
\caption{~~~~~~~~~~~~~~~~~~~~~~~~~~~~~~~~~~~~~~~~~~~~~~~~~~~~~~~~~~~Observations\label{obstab}} 
\vspace{0.25cm}
\begin{tabular}{llccl} \tableline\tableline 
{Object} & {} & {$\alpha$(B1950)} & {$\delta$(B1950)} & {AOTs \& TDTs} \\ \tableline 
IRAS\,18162-2048 & ON & 18 16 12.9 & $-$20 48 49 & LWS01-32901362 \\
           & OFF\,S & 18 16 12.9 & $-$20 50 29 & LWS01-32901362 \\
           & OFF\,N & 18 16 12.9 & $-$20 47 09 & LWS01-32901362 \\ 
HH\,80     &        & 18 16 06.8 & $-$20 53 06 & LWS01-32901350, SWS01-32901351 \\
HH\,81     &        & 18 16 07.4 & $-$20 52 23 & LWS01-32901360, SWS01-32901361 \\ 
HH\,80N    &        & 18 16 20.7 & $-$20 42 53 & LWS01-32901363 \\
\tableline
\end{tabular} \\
\end{center}
\end{table*}

We used the two spectrometers on the ISO satellite to observe  several 
locations along the HH\,80/81 flow, including the candidate exciting 
source IRAS\,18162-2048. The Long Wavelength Spectrometer (LWS, Clegg 
et al.~\cite{Clegg96}) was used in its LWS01 grating mode to acquire 
full low resolution (R$\sim$ 200) 43-197\um\ scans. Data were collected 
every 1/4 of a resolution element (equivalent to $\sim$0.07\um\ for 
$\lambda \lsim$90\um, and to $\sim$0.15\um\ for $\lambda \gsim$90\um) 
with 0.5s integration time; a total of 24 scans were accumulated, 
corresponding to 48s integration time per spectral element. LWS data 
processed through Off-Line Processing (OLP), version 7, have been 
reduced using the LWS Interactive Analysis\footnote{LIA is available at 
http://www.ipac.caltech.edu/iso/lws/lia/lia.html} (LIA) Version 7.2. The 
dark current and gain for each detector were re-estimated, and the data 
were recalibrated in wavelength, bandpass and flux. The absolute flux  
calibration for LWS in grating mode is 10\% (ISO Handbook, {\sc iv}, 
4.3.2). 
 
The Short Wavelength Spectrometer  (SWS, de Graauw et 
al.~\cite{DGetal96}) was used in its SWS01 ``Speed 2'' grating mode. In 
this mode the grating is moved faster than the detector reset time, with
a loss of about a factor 7 (ISO Handbook, {\sc vi} 3.3) in  achievable
spectral resolution, resulting in R$\sim300$. The total observing time
was about  1900s. SWS data were processed using OSIA, the SWS
Interactive  Analysis\footnote{OSIA is available at 
http://www.mpe.mpg.de/www\_ir/ISO/observer/osia/osia.html }. Dark 
currents and photometric checks were revised. The March 1998 bandpass 
calibration files have been used to produce the final spectra. The 
accuracy of the SWS absolute flux calibration varies between 7\% and 
35\% from 2 to 40\um\ (ISO Handbook, {\sc vi} 5.4.2). 
 
The final steps of data analysis were done using the ISO  Spectral 
Analysis Package\footnote{ISAP is available at 
http://www.ipac.caltech.edu/iso/isap/isap.html} (ISAP)  Version 1.6a
for  both LWS and SWS. Grating scans (LWS) and detectors spectra (SWS)
were  averaged using a median clipping algorithm optimised to flag and
discard  outliers mainly due to transients. The LWS spectra were heavily
fringed,  and standard techniques available under ISAP were used to
remove these  instrumental effects. Line fluxes were estimated by means
of gaussian   fitting. 

\begin{figure*}
\vspace{10cm}
\includegraphics{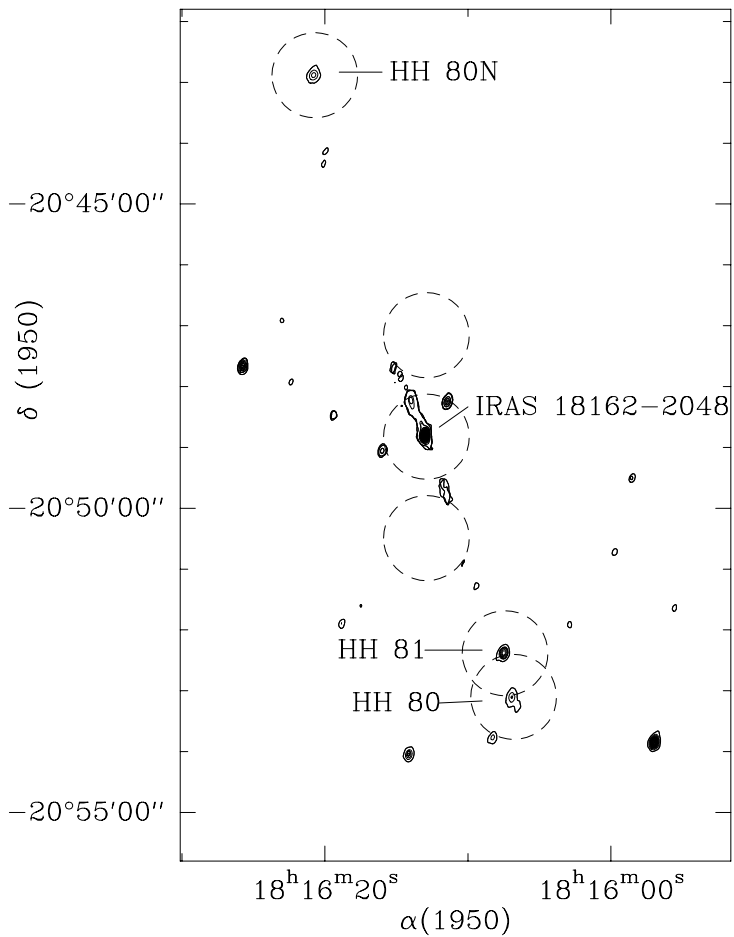}
\vspace{2cm}
\caption{\label{hh8081} VLA 6 cm composite map of the  HH\,80/81 region
(Mart\'{\i}, Rodr\'{\i}guez \& Reipurth (1993),  courtesy of L.F.
Rodr\'{\i}guez); superimposed is the ISO-LWS field of  view (dashed
circles) at all observed positions. The various sources in  the field
are labeled.} 
\end{figure*}

Table~\ref{obstab} lists the observed positions, which are also 
reported  in Fig.~\ref{hh8081} as the centers of the dashed circles 
(the LWS FWHM beamsize) superimposed on the 6~cm VLA map of MRR93). The
last column reports the Astronomical Observation Template (AOT) used,
together with the 8-digits unique identifier of the observation (TDT).

\section{Results} 
\label{res} 
 
All objects were detected with the LWS, while no line or continuum is
detected with the SWS toward HH\,80 and 81. Table~\ref{restab} lists the
measured fluxes for the detected lines; upper  limits are given at the
1$\sigma$ level. The distance between HH\,80 and 81 is about one half 
of the LWS beamsize ($\sim$80\asec\ FWHM at most wavelengths, see ISO
Handbook, {\sc iv} 4.10.1) implying that each object contributes roughly
half  of its flux (lines and continuum) to the observed flux of its 
neighbour.  Contamination corrected fluxes for HH\,80 and 81 are listed 
just below the observed ones for each line. 
 
\begin{table*}
\begin{center}
\caption{~~~~~~~~~~~~~~~~~~~~~~~~~~~~~~~~~~~~~~~~~~~~~~~~~~~~~Observed Line Fluxes\tablenotemark{a}\label{restab}} 
\vspace{0.25cm}
\begin{tabular}{lcccrrr} \tableline\tableline 
{Line} & {} & {IRAS\,18162-2048} & {} & {HH\,80} & {HH\,81} & {HH\,80N} \\ 
{} & {ON} & {OFF\,S} & {OFF\,N} & {} & {} & {} \\ \tableline 
\oia & 104(2) & 7.1(0.5) & 8.0(0.3) & 6.4(0.2) & 7.1(1.0) & 8.2(0.2)  \\   & & & & 3.8\tablenotemark{b} & 5.2\tablenotemark{b} & \\
\oib & 10.6(0.7) & 0.3(0.1) & 1.0(0.2) & 0.4(0.1) & 0.32(0.04) & 0.7(0.2) \\
  & & & & 0.3\tablenotemark{b} & 0.16\tablenotemark{b} & \\
\oiii &$\leq$4 &$\leq$0.3 &$\leq$0.4 & 0.8(0.4) & 1.0(0.4) & 1.0(0.2) \\
  & & & & 0.4\tablenotemark{b} & 0.8\tablenotemark{b} & \\
\nii &$\leq$1&$\leq$0.3&$\leq$0.2& 0.35(0.08) & 0.53(0.07) & 0.7(0.1) \\   & & & & 0.11\tablenotemark{b} & 0.47\tablenotemark{b} & \\
\cii & 37(1)& 16.3(0.3) & 20.4(0.3) & 11.49(0.06)& 12.53(0.08)& 17.1(0.1) \\
  & & & & 6.97\tablenotemark{b} & 9.04\tablenotemark{b} & \\
\tableline
\end{tabular} \\
\vspace{-1cm}
\tablenotetext{}{~~~~~~~$^a$ Units of 10$^{-19}$ \lflux\ and 1$\sigma$ statistical uncertainties in parenthesis; upper limits are at the 1$\sigma$ level.} 
\tablenotetext{}{~~~~~~~$^b$ Contamination-corrected fluxes.} 
\end{center}
\end{table*} 
 
\begin{figure*}
\vspace{10cm}
\includegraphics{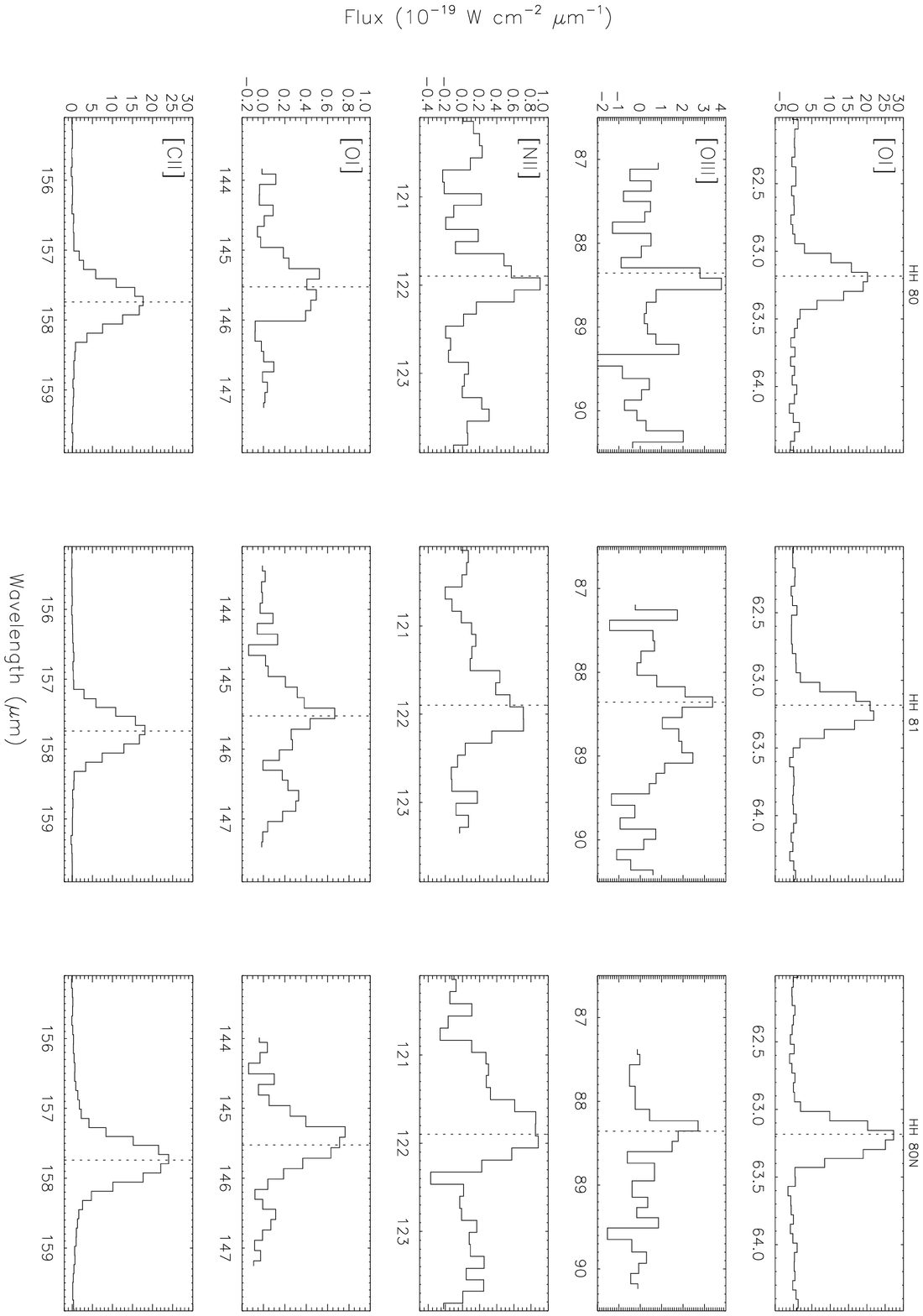}
\vspace{2cm}
\caption{\label{fig_lines_hh} Detected lines toward HH\,80, 81 and 80N.
Flux densities are normalised to 10$^{-19}$ \fdens. The dotted vertical
lines represent the expected linecenter wavelength.}
\end{figure*}

\begin{figure*}
\vspace{6cm}
\includegraphics{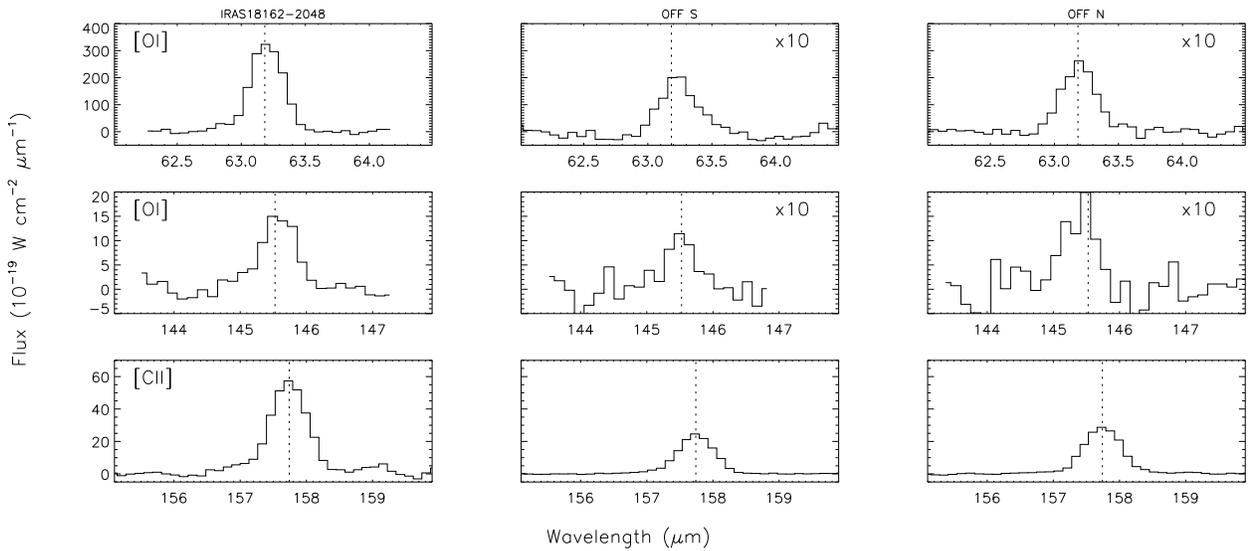}
\vspace{2cm}
\caption{\label{fig_lines_jet} Detected lines toward IRAS\,18162-2048,
OFF\,S and OFF\,N. Flux densities are normalised to 10$^{-19}$ \fdens.
The dotted vertical lines represent the expected linecenter wavelength}
\end{figure*}

A glance at Table \ref{restab} is sufficient to identify the basic 
results of our observations and set the guidelines for the 
interpretation of our data. First of all we note that the \nii\ and 
\oiii\ lines are only detected toward the HH objects; these observations
provide the first direct evidence for the collisionally excited HH
nature of HH\,80N. Although the LWS pointing toward  IRAS\,18162-2048
encompasses the radio-brightest portions of the  thermal jet, no lines
from ionised species (other than \cii, but see  below) are detected. We
verified that  the \nii\ and \oiii\ lines  detected toward the HH
objects would still be detectable toward  IRAS\,18162-2048, in spite of
the higher photon noise from this more  intense continuum source. 
Secondly, molecular emission is not detected  anywhere.  Finally we note
that with the exception of  IRAS\,18162-2048, the \cii\ line is
everywhere brighter than the \oia\  line. The detected lines are
reported in Fig.~\ref{fig_lines_hh} for the HH objects pointings, and in
Fig.~\ref{fig_lines_jet} for the three central observed positions of the
flow.

\subsection{The Herbig-Haro Objects} 
\label{hh} 

\begin{figure*}
\vspace{6.5cm}
\includegraphics{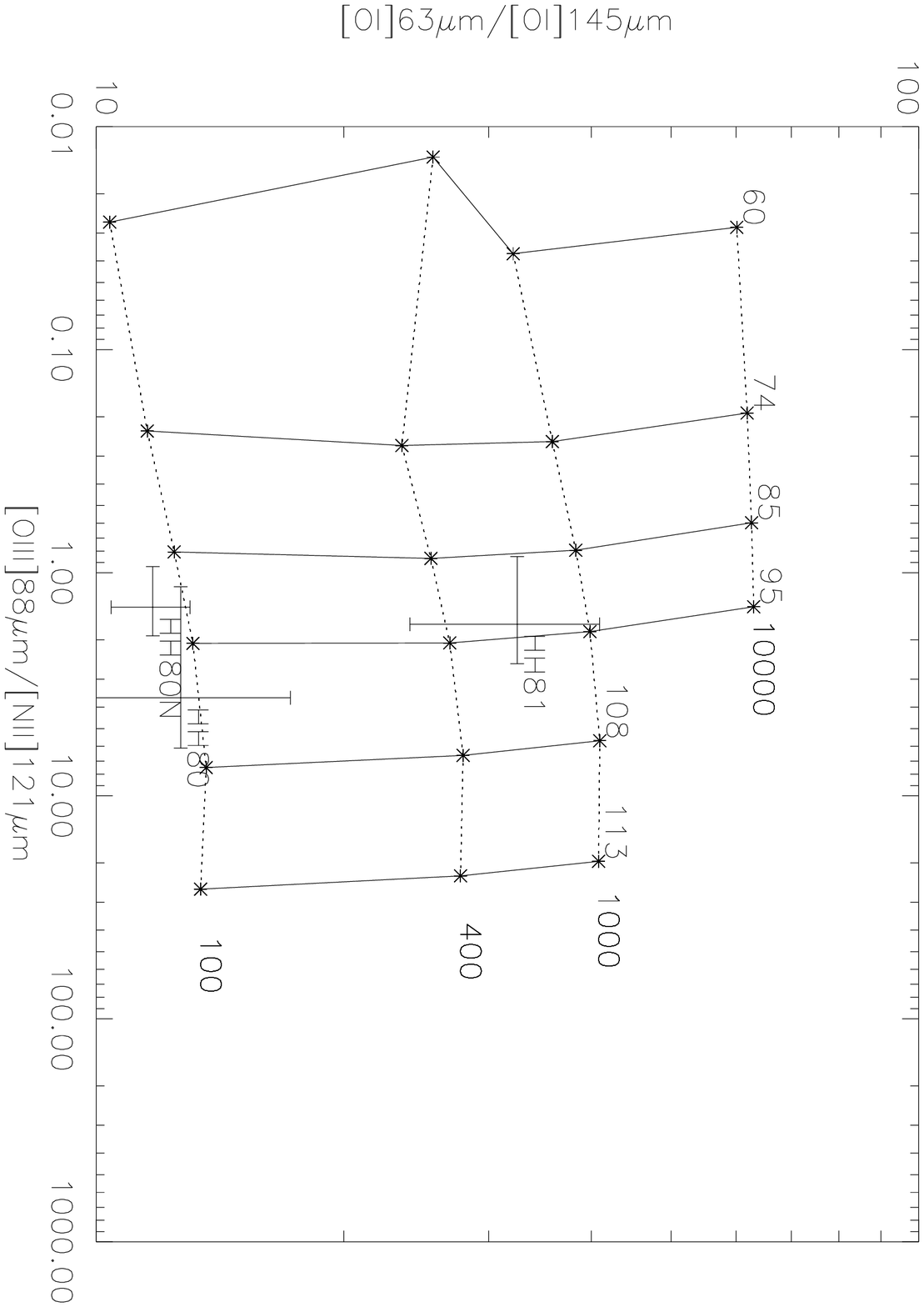}
\vspace{2cm}
\caption{\label{diag} \oia/\oib\ $vs$ \oiii/\nii\ diagnostic  diagram
obtained from the models of Binette et al. 1985 . Horizontal  lines are
of constant pre-shock density (in \cmthree), while vertical  lines are
of constant shock velocity (in \kms).} 
\end{figure*}

The HH\,80/81 system is the most powerful flow known to emanate  from an
intermediate mass Young Stellar Object (YSO).  Velocities in excess of
600 \kms\ are derived from proper motion  measurements both in radio
continuum (Mart\'{\i}, Rodr\'{\i}guez \&  Reipurth~\cite{MRR95},
\cite{MRR98}) and optical lines (HRR). In  addition, line profile
analysis of high-dispersion optical spectra  suggests shock velocities
$v_s\sim$650 \kms\ (HRR). Interestingly enough  however, optical spectra
only reveal lines from species up to double  ionisation state. Our FIR
spectroscopy has the immediate advantage that  reddening corrections,
which are critical in the optical regime  especially in this highly
extincted region (HRR), are not important. To  diagnose the shock
conditions toward HH\,80, 81 and 80N, we compare our  observations with
the predictions from plane parallel atomic shock models. The shock
models were calculated using MAPPINGS2, a code developed by  Binette et
al. (Binette, Dopita, \& Tuohy \cite{BDT85}), which has been  thoroughly
tested (Pequignot \cite{pequi86}).

A grid of models was created changing the pre-shock density between
10$^2$ and 10$^4$ \cmthree, consistent with values derived by HRR from
optical lines analysis. Shock  velocity were varied between 60 and 110
\kms; higher values, although still compatible with results from optical
line ratios (HRR), fail to produce lines like \nii\ and \oiii\, which
are instead observed toward the HH objects. The preshock degree of
ionisation and   magnetic field strength were kept fixed at 0.1 and
10\mug\ respectively.  These values are not critical for the velocities
we used (but see HRR $\S$ 8.2) and represent typical values for stellar 
jets (but see Bacciotti \& Eisl\"offel \cite{BE99}) from low-mass YSOs;
we will assume they are also valid for higher-mass systems like the one
we are presently delaing with. This selection of initial values also
allow us to gauge our models   with those of Hartigan, Raymond and
Hartmann (\cite{HRH87}). In all instances solar abundances were used,
since this is a reasonable assumption for Herbig-Haro objects
(Beck-Winchatz, B\"ohm,  Noriega-Crespo \cite{BBNC96}). Fig.~\ref{diag}
clearly show that shock velocities between 90 and 110  \kms\ are
appropriate to reproduce the emitted spectra for all HH objects.
Pre-shock densities between 100 and 1000 \cmthree\ are derived, in full 
agreement with estimates from HRR. In these conditions of low density 
J-shocks (Draine~\cite{D80}) negligible cooling is expected from
molecular  lines like CO, H$_2$ and H$_2$O (Hollenbach \&
McKee~\cite{HM89}), in  agreement with our observations. Furthermore,
the absence of FIR  molecular emission also rules out the presence of a
significant C-type  shock component, for which H$_2$, CO and H$_2$O
molecular lines would be the main coolants (e.g., Kaufman \&
Neufeld~\cite{KN96}).

These low velocity shocks plus the width of the observed optical lines
suggested to HHR that the HH\,80/81 objects correspond to ``wings'' of a
highly adiabatic bow shock that strikes the surrounding gas at $\sim
650$\kms; in such wings the shock front would be oblique with respect to
the direction of motion, resulting in lower shock velocities.  A shock
velocity of $\sim 100$ \kms~and flow velocities  (from the proper
motions) as high as 600 \kms~can be also reconciled if one considers
that the mass loss from the protostar is time dependent, and that the 
observed configuration of HH\,80/81 was preceded by previous ejection
events. If this is the case, the new ejected gas finds the circumstellar
gas already in motion, and it is the interaction between these two flows
that can create relative shock velocities of $\sim 100$ \kms~or so (Raga
et al. \cite{raga90}; Stone \& Norman, \cite{SN93}). Multi-epoch VLA
observations (Mart\'{\i}, Rodr\'{\i}guez \& Reipurth~\cite{MRR95},
~\cite{MRR98}) of the jet's radio knots provide supporting evidence for
episodic mass loss from IRAS\,18162-2048.

A similar process has been invoked to understand the flow
characteristics of the HH \,1/2 system, where the proper motions are as
high as 450\kms~ in both atomic and molecular H$_2$ gas (Herbig \& Jones
\cite{HJ81}; Rodr\'{\i}guez et al. \cite{rodri90}; Eisl\"offel, Mundt \&
B\"ohm  \cite{eis94}; Noriega-Crespo et al. \cite{nori97}), but the
shock themselves are $\sim 150-200$ \kms~(Hartigan et al. \cite{HRH87};
Noriega-Crespo,  B\"ohm \& Raga \cite{nori89}). In the HH\,1/2 outflow,
however, there is clear evidence of a previous outburst event marked by
the presence of an older bow shock structure  $\sim 10$ times farther
away from central outflow source  (Ogura \cite{ogura95}).

\subsection{The Jet} 
\label{jet} 
 
The LWS pointing toward IRAS\,18162-2048 encompasses the base of the
radio  thermal jet, which is the strongest radio emitter in the flow.
Yet  no  \oiii\ or \nii\ emission is detected, indicating that the
ionised  material of the jet is at a lower temperature compared to
HH\,80, 81 and 80N, or that the ionisation does not result from shocks.
Interestingly, the radio spectral index of the jet is  consistent with
that of  an  ionised wind, but significantly differs from that of the HH
objects which  instead manifest a possible synchrotron component (MRR93).
This behavior is different from that of the HH\,1/2 system, for
instance, where the central source VLA 1 (Pravdo et al.~\cite{Petal85})
has a similar positive radio spectral index to that of HH\,80/81 jet,
but for the HH objects themselves it is flat, indicating  optically thin
free-free emission (Rodr\'{\i}guez et al. \cite{rodri90}).
 
\subsection{The PDR} 
\label{pdr} 
 
\begin{figure*}
\vspace{7cm}
\includegraphics{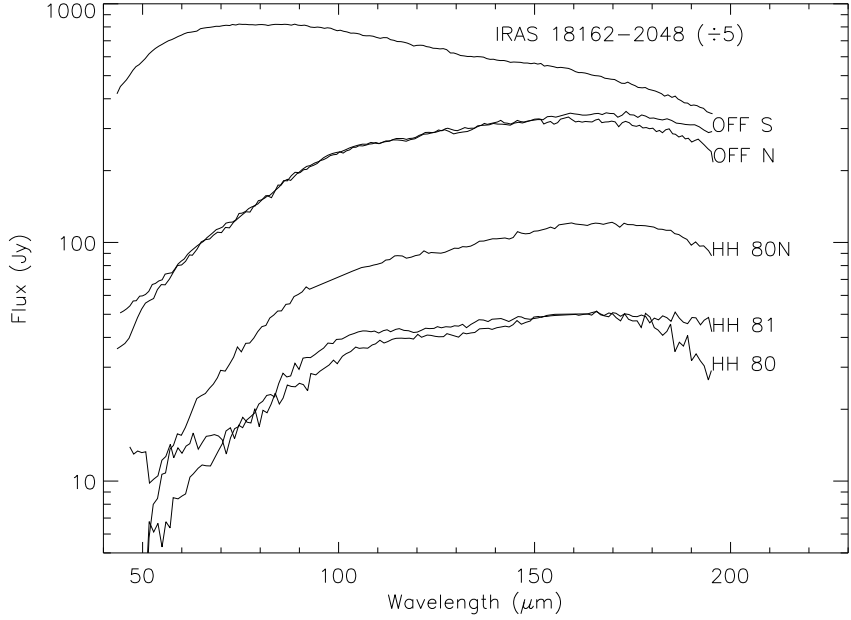}
\vspace{2cm}
\caption{\label{continua} Spectral energy distributions  observed with
LWS toward all pointings. The SED from IRAS 18162-2048 has been divided
by 5 to fit it into the plot range. The SEDs of HH\,80 and 81 have been
corrected for reciprocal contamination.} 
\end{figure*}

\begin{figure*}[b]
\vspace{6cm}
\includegraphics{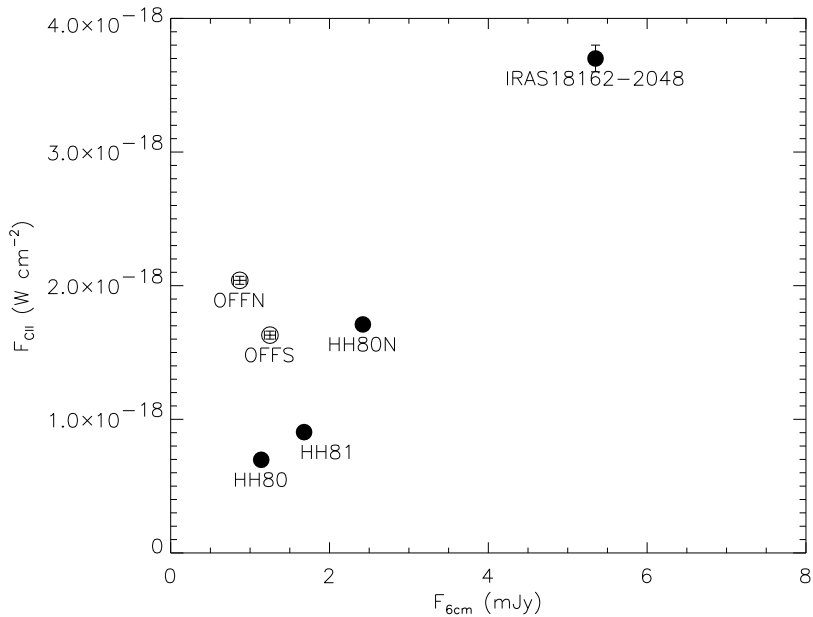}
\vspace{2cm}
\caption{\label{fcii_f6} Plot of the \cii\ line flux $vs$  the 6 cm
continuum flux from the knots of the jet encompassed by the LWS  beam.
When errorbars are not visible is because they are smaller  than the
symbols.} 
\end{figure*}

In the shock conditions diagnosed by our FIR spectroscopy (see 
Sect.~\ref{hh}), \cii\ contributes about 20\% of the \oia\ cooling. 
Since \cii\ is everywhere brighter than \oia\ (except toward 
IRAS\,18162-2048) the \cii\ emission must have another origin,  namely
PDR. Evidence for a PDR-like emission is also offered by the FIR 
continua detected toward the HH object and the OFF positions (the
latter  cannot be entirely justified with contamination by the stronger
continuum source  IRAS\,18162-2048). Fig.~\ref{continua} shows the SEDs
observed toward all pointed positions with the LWS. We note that the
continuum observed toward the two OFF  positions cannot be justified
with contamination from IRAS\,18162-2048. The continua from HH\,80 and
81 have been corrected for reciprocal contamination (see enf of
\S\ref{obs}), and detector spectra have been stitched for cosmetic
purposes.
 
In a PDR, electrons are released in the gas phase by FUV-irradiated dust
grains via the photoelectric effect and heat the gas (Tielens \&
Hollenbach~\cite{TH85}). Theory and observations generally agree on the
fact that between 0.1\% and 1\% of the incident FUV flux is converted
into gas heating via this mechanism, and subsequently released mainly
via ``cooling'' of the \cii\ line. The rest of the incident field is 
absorbed by dust and reprocessed to FIR wavelengths. We see from
Fig.~\ref{continua} that the observed SEDs contain the bulk of FIR
emission,  so that their integral is a good representation of the
reprocessed FUV  field; the ratio of observed \cii\ to the integrated
SEDs yields values  0.002$\leq \chi_c \leq$0.013, in very good agreement
with expectations for a PDR. The only obvious exception is
IRAS\,18162-2048, whose SED is dominated by  radiation from the central
YSO.  
 
Assuming that the pre-shock densities diagnosed for the HH objects 
(Sect.~\ref{hh}) represent the average conditions surrounding the 
HH\,80/81 flow, we can use the observed \cii\ flux to estimate the 
intensity of the irradiating FUV field. Using the Web Infrared Tool
Shed  (Wolfire et al.) and assuming complete filling of the LWS beam
with filling factor of unity we  estimate \go\ $\sim$200, 400, 30\,000,
200, 40 and 30 for the 6 LWS pointings from north to south (see
Fig.~\ref{hh8081}), where \go\ is the FUV field intensity expressed in
units of 1.6$\times$10$^{-3}$ ergs cm$^{-2}$ s$^{-1}$
(Habing~\cite{H68}). In these  conditions the PDR emission could account
for most of the \oia\  and \oib\ fluxes everywhere along the flow,
including the HH objects. In this case the \oia/\oib\ ratios would imply
densities $n<10^4$ \cmthree.
 
We searched several degrees around the HH\,80/81 area for OB stars
which  could be responsible for this irradiation level, finding none.
An  obvious candidate is of course IRAS\,18162-2048 itself. Assuming a
B0  ZAMS spectral class (MRR93), the resulting FUV field at the various
LWS  pointings would be \go\ $\sim$450 at OFF-N and OFF-S, and $\sim$60
at the  HH objects; the agreement with the \go\ values estimated from
the \cii\  line can be considered satisfactory. K-band images of
IRAS\,18162-2048,  however, clearly resolve the IRAS source in a small
cluster of at least 3 sources (Aspin et al.~\cite{aspin94});
redistribution of the total bolometric luminosity of the IRAS source
among the cluster members, adopting the IMF from Miller \& Scalo
(\cite{MS79}), would assign only half of this luminosity to the most
massive member of the cloud (e.g., Molinari et al.~\cite{Metal00}, and
references therein). This decreases the Lyman continuum photon flux by
more than one order of magnitude and reduces \go\ by nearly two orders
of magnitude, ruling out IRAS\,18162-2048 as the PDR illuminating
source.

An intriguing possibility is that the illuminating sources for the
observed PDR emission are the HH objects and the jet. In
Fig.~\ref{fcii_f6} we report the \cii\ line  flux as a function of the 6
cm flux measured by MRR93. The filled symbols represent the HH objects
and IRAS\,18162-2048; for each object  we consider the radio knots
encompassed by the LWS beam at each  position. The  correlation between 
F$_{C{\sc ii}}$ and F$_{6cm}$ would seem to suggest that the ionised
material and the amount of cooling by the PDR maybe somehow related. We
note that the two OFF positions do not fit with the correlation in
Fig.~\ref{fcii_f6}. In the next paragraph we will propose a
semi-empirical model to verify and quantify the Radio-PDR connection.

\section{A Model for the Shock/Jet$-$PDR Connection}
\label{model}

The influence of shock-originated radiation on the surrounding medium in
HH flows is not clear. NH$_3$ and HCO$^+$ enhancements have been
detected in the vicinity of HH\,80N, HH\,1 and HH\,2 (Girart et
al.~\cite{Getal94},~\cite{Getal98}; Davis, Dent \& Bell
Burnell~\cite{Detal90}; Torrelles et al.~\cite{Tetal92},
~\cite{Tetal93}) and it has been suggested that the UV field generated
in the HH shocks may be responsible. Wolfire \& K\"onigl (\cite{WK93}),
and more recently Raga \& Williams (\cite{RW00}), have shown that the
chemistry of blobs in the vicinity of HH objects can be influenced by
the passage of the HH object, although the predicted morphologies do not
match the observations.

We propose a scenario where the ionised material which is recombining
behind the shock front and in the jet, and which emits free-free radio
continuum, is also responsible for a UV field which illuminates
internally the walls of the flow cavity; a PDR is there produced, which
mainly cools via the \cii\ line. If this is correct, then we should be
able to express the two observables, i.e. the radio continuum flux and
the \cii\ line flux, as a function of a set of physical parameters which
characterize the HH shocks, the jet and the PDR.

Radio emission from shocked material has been modeled by Curiel, Cant\'o
and Rodr\'{\i}guez (\cite{CCR87}); in the optically thin regime, the
free-free emission from the recombination region can be written as

\begin{eqnarray}
S_{sh_{\nu}} & = & 1.84\,10^{-4} ~\theta^2~ \left[{{\nu}\over {10~
GHz}}\right]^{-0.1}~ T_4^{0.45}~ n_{o_{10}} \nonumber \\  
& & v_{s_7} [1+3.483 v_{s_7} -2.745]~{\rm mJy}
\label{radioshock}
\end{eqnarray}

where $\theta$ is the angular diameter in seconds of arc of the
recombination region, T$_4$ is the electrons temperature in units of
10$^4$K, $n_{o_{10}}$ is the pre-shock density in units of 10 \cmthree,
and $v_{s_7}$ is the shock velocity in units of 100 \kms. This model
applies to HH\,80, 80N and 81, whose high ionization lines clearly trace
a high velocity shock. In the absence of similar evidence toward the
central source and the two OFF positions (Tab.~\ref{restab}), where the
LWS beam encompasses most of the jet radio knots, we will assume that the ionisation does not results from shocks. Reynolds (\cite{R86}) developed a model to predict the properties of radio continuum emission from a collimated, ionised thermal jet. His expression of the radio flux at any specific frequency is a complicated function of jet parameters like the collimation, the density, temperature and ionization radial gradients, the initial distance where the jet is injected, the jet's width at its base. Because we cannot independently fix any of these parameters, the diagnostic power of this model will be very limited in the present case. 

Let us now quantify the energy released by the shock recombination
region, in the portion of the UV continuum, between 912\AA\ and 2066 \AA
(or between 13.6eV and 6eV), which is effective in PDR illumination. It is widely accepted (Dopita, Binette \& Schwartz~\cite{DBS82}) that the dominant contribution to UV continuum in this wavelength range comes from hydrogenic 2s$\rightarrow$1s two-photon decay.
%
%
%
%
Theoretical models (Shull \& McKee \cite{SM79}, also confirmed by
our grid of computed shock models (\S \ref{hh})) suggest that two-photon
emission from He$^{\rm 0}$ and He$^+$ can be neglected in comparison. The
H$^{\rm 0}$ two-photon emissivity is given by 

\begin{equation}
j_{\nu}={{1}\over {4\pi}}{{h\nu}\over {\nu_0}}P_{\nu/\nu_0}
N_{2s}~{\rm erg ~s^{-1} ~sr^{-1} ~cm^{-3} ~Hz^{-1}}
\label{jnu2p}
\end{equation}

where $\nu_0$ is the Ly$\alpha$ frequency. P$_{\nu/\nu_0}$ is the
probability, symmetrical around $\frac{1}{2}\nu_0$, that a photon is emitted with
frequency $\nu/\nu_0$ and for which we adopted the analytical
approximation of Nussbaumer \& Schmutz (\cite{NS84}). N$_{2s}$ is the
population density of level 2s and can be found equating the
recombination rate from higher states to 2s, with the total
2s$\rightarrow$1s two-photon decay rate. We obtain (Emerson \cite{E96}):

\begin{equation}
N_{2s}\sim 0.3 ~\alpha^{\prime} _{rec} N_e N_i
A_{2q}^{-1}~{\rm cm^{-3}}
\label{n2s}
\end{equation}

\begin{figure*}[t]
\vspace{10cm}
\includegraphics{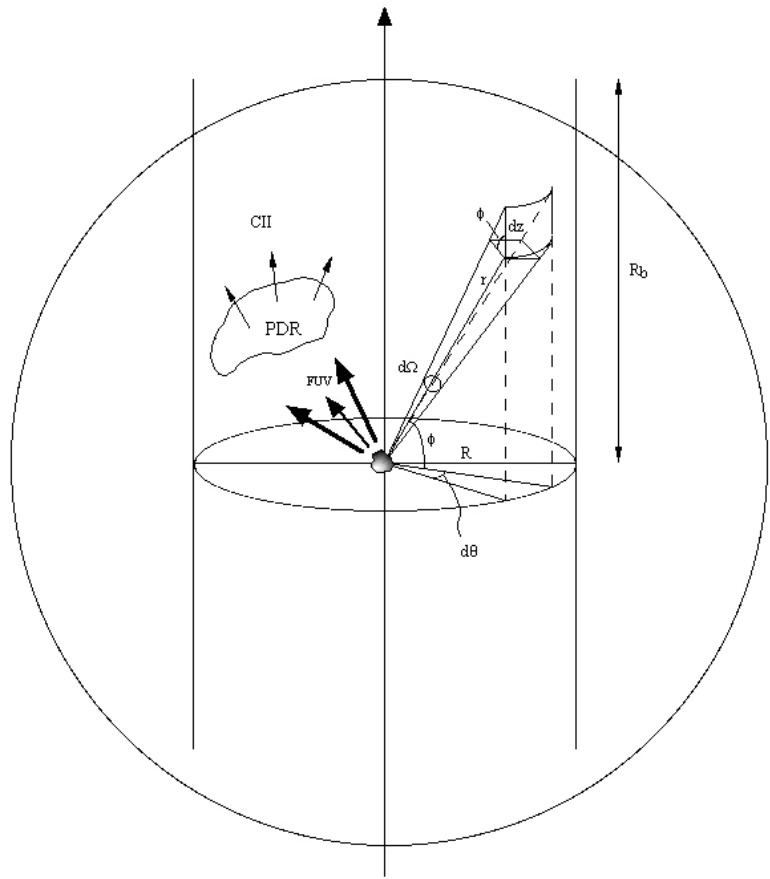}
\vspace{2cm}
\caption{\label{cavity_8081} Sketch of the proposed scenario for the
HH-PDR connection. The HH object is the grey knot at the origin of the
coordinate system, and the cylinder represents the cavity excavated by
the flow. The portion of big circle at the top of the figure indicates
the ISO-LWS beam.}
\end{figure*}

where N$_e$ and N$_i$ are the electron and ion number densities, and
A$_{2q}\sim 8.23$ s$^{-1}$ is the two-photon transition probability. The
total recombination coefficient to all hydrogen excited levels can be
written as (Hummer \& Seaton~\cite{HS63}):

\begin{eqnarray}
\alpha^{\prime} _{rec} & = & 1.627\,10^{-13} ~T_4^{-1/2}~ [1-1.657~
Log_{10}T_4~+ \nonumber \\ 
 & & 0.584\,T_4^{1/3}~]~{\rm cm^3 ~s^{-1}}
\label{alfarec}
\end{eqnarray}

where T$_4$ is T/10$^4$ K. Using Eqs. (\ref{jnu2p}), (\ref{n2s}) and
(\ref{alfarec}), we are able to write the power radiated by the
recombination region in the FUV as:

\begin{equation}
L_{FUV} = 4\pi ~V_{rec} ~\eta_{ce} \int_{6eV}^{13.6eV} j_{\nu}
d\nu~{\rm erg~s^{-1}}
\label{lfuv}
\end{equation}

V$_{rec}$ is the volume of the recombination region, while $\eta_{ce}$
is a factor which accounts for a collisional enhancement of the 2s level
population above the values predicted by pure recombination. Such an
enhancement can be determined from a comparison of the predicted
two-photon spectrum with the observed UV continuum from HH objects.
Dopita, Binette \& Schwartz (\cite{DBS82}) find values
$2.8\leq\eta_{ce}\leq13.3$; interestingly, $\eta_{ce}$ is found to be
inversely proportional to the degree of excitation of the HH object as
measured, e.g., from the \oiiib/\oic\ ratio. Eq.~(\ref{lfuv}) makes the
implicit assumption that the ionised clump is optically thin to the FUV
radiation. It is easy to show that for typical dust grains of radius
0.1\um\ and density of 3 g\cmthree, the optical depth along the diameter
of a spherical clump of radius $r$ and particle density $n$ can be
written as:

\begin{equation}
\tau_{\nu}= 1.8\,10^{-8} Q_{\nu} n\, r\, d
\label{tau}
\end{equation}

where $n$ is in \cmthree, $r$ is in seconds of arc and the distance $d$
is in parsecs. In the FUV range the absorption coefficient Q$_{\nu}\sim
1$ (Draine \& Lee \cite{DL84}), and for the typical parameters that we
will derive (see Table~\ref{resmodelhh}) we obtain $\tau_{FUV}\sim0.8$.
We emphasize that this number refers to the longest path across the
clump, so only a small portion at the far side of the clump, with
respect to any line of sight, will be only partially thick.

\begin{figure*}[b]
\vspace{7cm}
\includegraphics{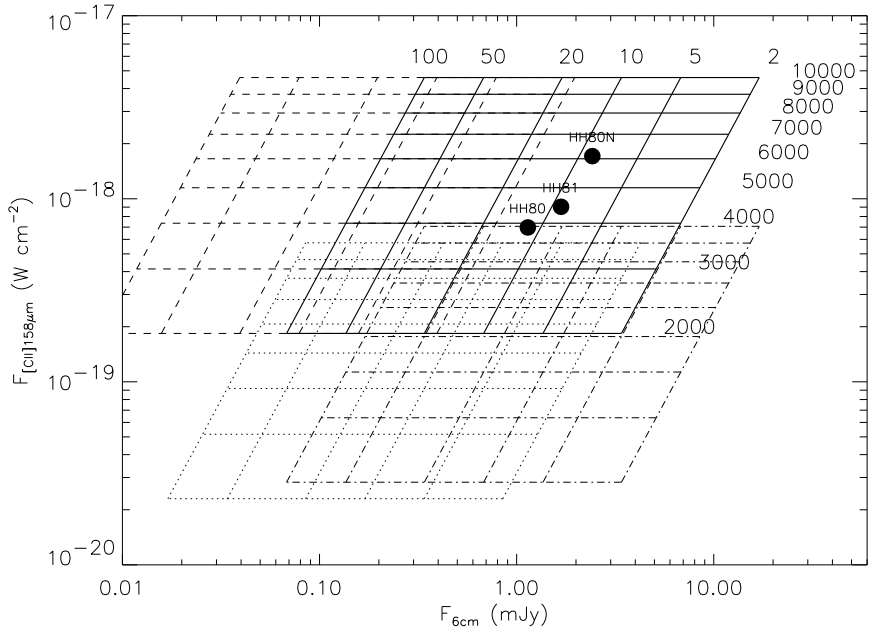}
\vspace{2cm}
\caption{\label{diag1} \cii\ line flux as a function of 6 cm radio
continuum flux (from Eq.~\ref{radioshock}) for different choices of HH
model parameters. Each grid shows the locus of points of constant
density $n$ (horizontal lines) and constant $\mathcal{R}$=n/n$_0$
compression ratio. The full-line grid is computed for $v_s$=100 \kms,
$\eta_{ce}$=2, $\chi_c=0.007$ and r$_{HH}$=5\asec. The other grids have
$v_s$=70\kms\ (dashed), $\chi_c=0.002$ (dash-dotted) and
r$_{HH}$=2.5\asec\ (dotted). The full circles represent the
observations.}
\end{figure*}

The FUV field emitted by the HH object is intercepted by the flow cavity
walls, and a PDR is there formed. A fraction $\chi_c$ of the incident
FUV field is predicted by PDR models (Tielens \& Hollenbach \cite{TH85})
to be reradiated via the \cii\ line; our observations allow us to
determine (\S~\ref{pdr}) values of $\chi_c$ in very good agreement with
model predictions. Since we want to compare the predicted \cii\ line
flux radiated by this PDR with our ISO-LWS observations, we are
interested in  the portion of the flow cavity which is encompassed by
the instrument field of view. In other words, we need to estimate the
fraction $f_c$ of the L$_{FUV}$ emitted by the HH object which is
intercepted by the flow cavity in our LWS beam. The situation is
sketched in Fig.~\ref{cavity_8081}, which serves as reference for the
following discussion. R is the radius of the cavity, and R$_b$ is the
radius of the ISO-LWS beam. For simplicity, we will assume the HH object
lying at the center of our coordinate system as pointlike with respect
to R. The element solid angle at the cavity wall as seen from the origin
is

\begin{equation}
d\Omega = {{R~d\theta ~dz~ \cos\,\phi}\over {r^2}}
\label{domega}
\end{equation}

where the cos\,$\phi$ factor accounts for the projection, perpendicular
to the line of sight from the origin, of the element area of the cavity
wall. Expressing $z$ and $r$ as functions of $R$ and $\phi$, we obtain

\begin{equation}
d\Omega = \cos\,\phi\,d\phi ~d\theta
\label{domega2}
\end{equation}

The total solid angle $\Omega_c$ under which the HH object sees the
internal cavity walls (from z=-R$_b$ to z=R$_b$) is obtained by
integrating Eq. (\ref{domega2}) over 2$\pi$ in d$\theta$, and from
-arctan(R$_b$/R) to arctan(R$_b$/R) in d$\phi$. The FUV field is
emitted isotropically, so that

\begin{equation}
f_c = {{\Omega_c}\over {4\pi}} = \sin \left[\arctan {{R_b}\over {R}}\right]
\label{fc}
\end{equation}

We can finally write the predicted flux of the \cii\ line from a system
at the distance D as:

\begin{equation}
F_{C{\sc ii}} = {{10^{-7}~L_{FUV}~f_c~\chi_c}\over {4\pi
D^2}}~{\rm W cm^{-2}}
\label{fcii}
\end{equation}

\section{Discussion}
\label{discussion}

\subsection{The Herbig-Haro Objects}

\begin{table*}[t]
\begin{center}
\caption{~~~~~~~~~~~~~~~~~~~~~~~~~~~~~~~~~~~~~~~~Model inputs and results for HH objects\label{resmodelhh}} 
\vspace{0.25cm}
\begin{tabular}{llcccccccc} \tableline\tableline 
{Object} & {} & {$\chi_c$} & {r$_{\rm HH}$} & {R} & {$v_s$} & {$\eta_{ce}$} & {FUV} & {n} & {$\mathcal{R}$} \\
{} & {} & {} & {(\asec)} & {(\asec)} & {(\kms)} & {} & { (\go)} &  {(\cmthree)} & {} \\ \tableline 
HH\,80  &             & 0.011 & 5 & 10 & 100 & 2 & 170 & 4200 & 13 \\
HH\,81  &             & 0.013 & 5 & 10 & 100 & 2 & 180 & 4500 & 10 \\
HH\,80N &             & 0.007 & 5 & 10 & 100 & 2 & 670 & 8400 & 13 \\ 
\tableline
\end{tabular} 
\end{center}
\end{table*} 

To understand which are the critical parameters in our model, it is
useful to show in Fig.~\ref{diag1} a diagnostic diagram that presents
the relationship between our two observable quantities, F$_{6cm}$ and
F$_{C{\sc ii}}$, for various parameters sets. We will first consider the
case of the HH objects, where Eq.~(\ref{radioshock}) holds for the
emitted radio continuum flux. For each grid in Fig.~\ref{diag1}, the
horizontal lines are sites of constant density ($n$ in \cmthree) while
the vertical lines indicates constant $\mathcal{R}$=$n/n_o$ compression
ratio (density over pre-shock density). The full-line grid is computed
for $v_s$=100 \kms, $\eta_{ce}$=2, $\chi_c$=0.007 and r$_{HH}$=5\asec.
Decreasing the shock velocity to 70 \kms shifts the grid to lower radio
fluxes (dashed), while a lower FUV$\rightarrow$C$_{\sc ii}$, $\chi_c$,
conversion factor brings it down in \cii\ flux (dash-dotted, the
opposite effect is obtained by raising the $\eta_{ce}$ enhancement
factor). Decreasing the radius of the HH object influences both
quantities (dotted). We find no appreciable effect from the particular
choice of the cavity radius R. It is reassuring that for reasonable
choices of the model parameters we can reproduce the 6cm and \cii\
fluxes observed toward the HH objects. The number of free parameters in
our model can be significantly decreased thanks to the available
observational evidence. The value of $\chi_c$ for each object has been
determined using our FIR continuum and \cii\ measurements
(\S.\ref{pdr}). The radius of the HH objects is taken from the radio
maps of MRR93; the objects are clearly resolved, and the full size
(after a tentative beam deconvolution) is $\sim$10\asec. The value of
$\eta_{ce}$ is computed using its relationship (Dopita, Binette \&
Schwartz~\cite{DBS82}) with the degree of excitation measured by the
\oiiib/\oic\ ratio; we use the \oiiib\ and \oic\ line fluxes from HRR to
derive values of 2 for the HH objects and 13 for the jet. We adopt shock
velocities of 100 \kms\ for all HH objects from the measured \oiii/\nii\
ratio. Model results are weakly dependent on the cavity radius, for
which we adopt a fiducial value of  R=10\asec. We can then vary the
plasma density and $\mathcal{R}$ to fit the data, and the  results are
in Cols. 8-10 of Table~\ref{resmodelhh}.

For all HH objects we obtain
densities which are below 10$^4$\cmthree, for which $2s\rightarrow 2p$
collisional population would depress the two-photon continuum due to
Ly$\alpha$ decay. The presence of a transverse (\bo\ $\perp$\vs)
magnetic field limits the compression ratio; we can write (Hollenbach \&
McKee~\cite{HM79})

\begin{equation}
B_{0\perp} = 76.7 \left[{{v_s}\over {100 {\rm km\,s^{-1}}}}\right] \sqrt{{n}\over {{\mathcal{R}}^3}}~\mu {\rm G}
\label{boperp}
\end{equation}

Using the parameters from Table \ref{resmodelhh} we see that \mbop\
varies from 100 to 200 \mug. We ran additional shock models using these
large \mbop\ values and found that the only way to reproduce the
observed line ratios is to assume a pre-shock ionization fraction
$\geq0.5$. This is higher than the higher values (0.2-0.3) found toward
several HH-jets by  Bacciotti \& Eisl\"{o}ffel (\cite{BE99}); on the
other hand the HH\,80/81 flow is the most powerful in its class and may
be peculiar in its ionization fraction as well. 

\subsection{The Jet}

As far as the other positions along the jet are concerned, our model has
less diagnostic power because of the high number of free parameters in 
the Reynolds (\cite{R86}) model for radio emission from ionised jets.

\begin{figure*}
\vspace{7cm}
\includegraphics{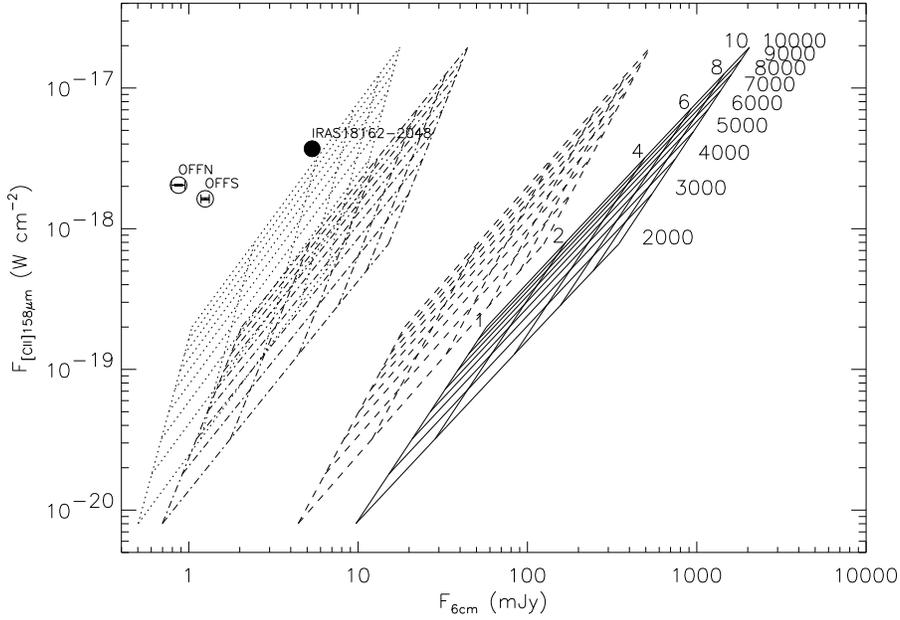}
\vspace{2cm}
\caption{\label{diag2} \cii\ line flux as a function of 6 cm radio
continuum flux (from Reynolds 1986) for different choices of jet model
parameters. Each grid shows the locus of points of constant density $n$
and jet's initial half-width in arcseconds. The full-line grid is
computed for [$\epsilon$, q$_n$, q$_T$, q$_x$]=[0.45, -0.9, -0.6, -0.5].
The other grids have q$_T$=-1 (dashed), $\epsilon=0.2$ (dash-dotted) and
q$_n$=-2 (dotted). The circles represent the observations; the open
symbols mean an uncertain assignment of the radio flux.}
\end{figure*}

Fig.~\ref{diag2} is the analogous for the jet of Fig.~\ref{diag1}, where
grids of models are plotted as a function of density $n$ in \cmthree\
and half-width w$_0$, in arcseconds, of the jets at its base. While
$\chi_c$ and $\eta_{ce}$ are kept fixed to 0.002 (from our observations)
and 13 (see above), the jet parameters are essentially free. The
full-line grid represent the ``pressure-confined'' jet in Reynolds
terminology, where the density, temperature and ionization radial
gradients of the jet have values [q$_n$, q$_T$, q$_x$]=[-0.9, -0.6,
-0.5]; in this particular jet the degree of collimation, expressed by a
number $\epsilon$ which controls the jet's width as a function of radial
distance (w$\propto r^\epsilon$), is $\epsilon=0.45$. Another critical
parameter, which we cannot independently fix, is the distance from the
central source where the jet is injected; the full-line grid in
Fig.~\ref{diag2} is obtained for r$_0$=10 \rsun. This model predicts too
much radio flux compared to the observations; other classes of standard
models in Reynolds study are even less successful. Although of uncertain
physical meaning, other combinations of the jet parameters can bring the
predicted radio fluxes closer to the values observed along the HH\,80/81
jet. Steepening the density gradient to q$_n$=-2 (dotted grid) will
decrease the radio flux by more than an order of magnitude; a less
pronounced effect in the same direction is obtained by  decreasing the
temperature gradient to, e.g., q$_T$=-1 (dashed grid). Providing
additional collimation by lowering $\epsilon$ down to, e.g., 0.2
(dashed-dotted grid), will also contribute to a better match between the
model and the data. Finally, decreasing the jet-injection distance from
the source, r$_0$, will also work toward lowering the predicted radio
flux. Although there is no single set of jet parameters that can fit the
data, our model suggests steep density and/or temperature radial
gradients along the jet, as well as a very high degree of collimation
which is not ruled out by the observations (MRR93). Looking at the shape
of the models grids in Fig.~\ref{diag2}, it seems  that a single jet
model, i.e. a single model grid, cannot reproduce the [F$_{C\sc{ii}}$,
F$_{6cm}$] values of IRAS\,18162-2048 and the two OFF positions. This is
not due to the misalignment between the observed OFF positions and the
jet axis. A better alignment would not have changed the radio flux
(which we arbitrarily assigned based on the 6 cm map of MRR93);
likewise, a half-LWS beam shift in the ISO observed positions would
certainly not result in a one order of magnitude decrease of the \cii\
line fluxes, as required to bring the OFF positions onto the model grid
in Fig.~\ref{diag2}. Rather, it is plausible that the central (brighter)
portion of the jet considerably contributes to the FUV irradiation of
the flow cavities at the two OFF positions. This would indeed correspond
to increasing the assigned radio flux to the two OFF positions in
Fig.~\ref{diag2}.

\subsection{The F$_{C\sc{ii}}$-F$_{6cm}$ relationship}

Although it was one of the motivations to develop our model, one of the
consequences of the model is that a F$_{C\sc{ii}}$-F$_{6cm}$
relationship is difficult to justify. The radio flux has a different
origin in the HH objects and the jet, there is {\it a priori} no reason
why IRAS\,18162-2048 (where we model the radio flux as coming from a
jet) should line up with the three HH objects (where we model the radio
flux as coming from the post-shock region) in Fig.~\ref{fcii_f6}. The
relative position of the HH objects themselves is also {\it a priori}
depending on a high number of free parameters. Based on four points
only, we must then conclude that the F$_{C\sc{ii}}$-F$_{6cm}$
relationship in Fig.~\ref{fcii_f6} appears to be fortuitous. However, we
will check for similar occurrences in other HH/jet systems where \cii\
data are available and where the densities are low enough ($n\leq 10^4$
\cmthree) that a shock/jet$\leftrightarrow$PDR connection can be
expected.

\section{Conclusions}
\label{conclusions}

We have performed a far-IR spectroscopic study of the HH\,80/81 system.
Line ratio analysis confirms for the first time the Herbig-Haro nature
of the nebulosity HH\,80N, which probably represents the head of the
counterflow to HH\,80 and 81. We reveal shock velocities of the order of
100 \kms\ in correspondence with the HH objects, while lower excitation
conditions appear to be present elsewhere along the radio jet. A
comparison with  proper motion velocities in excess of 600 \kms\
indicate that the shocks arise at the interface between two fast-moving
flows.

Besides shock-excited emission, an important PDR contribution is present
all along the bipolar flow, where densities below 10$^4$ \cmthree\ are
also diagnosed. Using a simple model, we have provided quantitative
arguments supporting the idea that the FUV field radiated by the ionised
material of the recombination regions in the HH objects and of the jet
emanating from IRAS\,18162-2048, is able to induce the formation of a
PDR in the immediately  surrounding medium (i.e. the flow cavity walls).
This would provide further evidence that the jet/HH own radiation field
affects its surrounding medium in a measurable way. Attempts to model
the chemistry of outflows should not ignore the influence of the
radiation field of the shocks responsible for the acceleration of the
outflow itself. 

\acknowledgements

We thank Luis Felipe Rodr\'{\i}guez for kindly making available to us
his 6 cm VLA map of the HH\,80/81 region. We also thank the referee,
Chris Davis, for his careful reading of the manuscript. The ISO Spectral
Analysis Package  (ISAP) is a joint development by the LWS and SWS
Instrument Teams and  Data Centers. Contributing institutes are Centre
d'Etude Spatiale des Rayonnements (France), Institute d'Astrophysique
Spatiale (France), Infrared Processing and Analysis Center (United
States), Max-Planck-Insitut f\"ur Extraterrestrische Physisk (Germany),
Rutherford Appleton Laboratories United Kingdom) and the Space Research
Organization, Netherlands.

{}

\end{document}